# Graphene synthesis by chemical vapor deposition and transfer by a roll-to-roll process

Zhen-Yu Juang[*], Chih-Yu Wu, Ang-Yu Lu, Ching-Yuan Su, Keh-Chyang Leou, Fu-Rong Chen, Chuen-Horng Tsai

*Department of engineering and System Science, National Tsing Hua University, Hsinchu, Taiwan 30013, ROC.*

*Corresponding author. E-mail: zyjuang@mx.nthu.edu.tw

## Abstract

We synthesized centimeter-scale single- to few-layer graphene (FLG) films via chemical vapor deposition (CVD) on Ni foils. We demonstrates that the precipitation mechanism may not be the only important mechanism in the formation of graphene by CVD in Ni system, and that controlling the cooling rate in the CVD process may not be the appropriate way to control the thickness of graphene films. In addition, we are the first to demonstrate the transfer of centimeter-scale FLG from Ni foil to transparent flexible polyethylene terephthalate substrates via an efficient roll-to-roll process. Comparing to rigid substrates, synthesis of graphene on flexible Ni foil has necessity for the use of a roll-to-roll transfer process.

# 1. Introduction

Graphene is a nearly transparent material that has the highest room temperature electrical conductivity of any known substance. [1] Thus, scientists and engineers predict that many future applications can be realized using graphene, such as ultrahigh-speed transistors [2] and transparent electrodes. [3, 4] Obtaining single-layer graphene (SLG) or few-layer graphene (FLG) via adhesive tape or mechanical cleavage methods is not practical, as these methods provide limited opportunity for scaling up. [2, 5] The ability to synthesize and place large-scaled graphene films on various substrates with methods compatible with current industrial technology is very important for future applications.

There has been a dramatic proliferation of research concerned with various methods for the synthesis of graphene since its discovery in 2004. [2] The use of chemical vapor deposition (CVD) and other surface precipitation methods for the synthesis of graphene on transition metals has recently been reported. [6-16] The advantage of these methods is that large graphene domains can be easily obtained. Most importantly, since some transition metals can be etched by acid solutions, graphene deposited on these materials can be easily transferred to other substrates. However, to date, there has been relatively little research conducted on the growth mechanism of graphene in a CVD process. Thickness control of graphene during synthesis is an important step for future applications, and understanding this mechanism is critical. Although some reports claim that the thickness can be controlled, [16] there remains a large degree of uncertainty.

More recently, there has been a shift in the research focus of the field from developing techniques for graphene synthesis to designing strategies for transferring graphene films to various substrates. [12-18] Although some transfer methods that have been presented offer valuable insights into future applications, they all include limitations. One of the thorniest problems facing researchers today is the question of how to scale-up the transferred yield and area of graphene films in a uniform and effective manner, using industry compatible methods. The roll-to-roll process may be an optimal solution, opening up new possibilities for future applications.

**2. Experimental details**

In this paper, we present the development of CVD synthesis and a roll-to-roll transfer process of graphene. The CVD synthesis method, using transition metals as substrates, is a promising technique for the synthesis of large-scaled graphene films. Before the roll-to-roll transferring process, SLG and FLG were synthesized using Ni foil (30 μm in thickness, The Nilaco Corp.) as the substrate. After the Ni substrates were loaded into a quartz tubular furnace, the samples were heated to the process temperature of 900 ºC and maintained for 10 min under a $H_2$/Ar (10 and 400 sccm, respectively) atmosphere. A $CH_4$/$H_2$/Ar gas mixture (10, 10 and 400 sccm, respectively) was then introduced into the quartz tube for 10 minutes, and the furnace was cooled to room temperature with fast (3 ºC/sec) and slow (0.3 ºC/sec) cooling rates. The gas mixture during cooling is the same as heating stage. The pressure was maintained at 750 Torr during whole process using advanced pressure control system (APC, MKS Instruments). An optical microscope (OM) and scanning electron microscope (SEM, JEOL JSM-6330F) were used to observe the sample surface. Raman spectroscopy (HORIBA Jobin Yvon HR800) was used to measure the bonding characteristics and estimate the thickness of the graphene films. The Raman spectra were obtained using a He-Ne laser with a wavelength of 632.8 nm and a spot size of ~ 1 μm. The sheet resistance was measured using the four-point probe method. Transmittance properties were measured using a wavelength of 550 nm.

## 3. Results and discussion

After the CVD process, the graphene films almost completely cover the surface of the Ni foil. Fig. 1 shows typical SEM and Raman images of the CVD-synthesized single- to few-layer graphene films. The films shown in these images were subjected to a cooling rate of 3 ºC/sec. In the SEM image (Fig. 1a1), wrinkles are visible in the graphene film (marked by a red arrow) which are a result of the difference in the thermal expansion between Ni and the as-grown graphene films during the cooling stage, consistent with previous reports. [19, 20] The Ni grain boundaries that can be seen underneath the graphene in the SEM image are marked by dotted lines. Fig. 1a shows the Raman spectra of positions A through D in the SEM image. The A and B positions refer to areas of SLG. The Raman spectra at positions C and D reveal the presence of FLG. [21-23] By tracking the 2D peak across our sample, we see that single- or few-layer regions can be up to several tens of μm in lateral size.

Note that the 2D-line positions around 2665 $cm^{-1}$ seem significant higher than exfoliated graphene on $SiO_2$/Si (~2640 $cm^{-1}$), and consists with other CVD synthesized graphene. [13, 19, 24] This blue-shift phenomenon might comes from the difference in the thermal expansion between Ni and the as-grown graphene films during the cooling stage, and store strain in graphene lattice. [25] In addition, the substrate effect infl uences the peak positions in Raman spectra as well. [26-28]

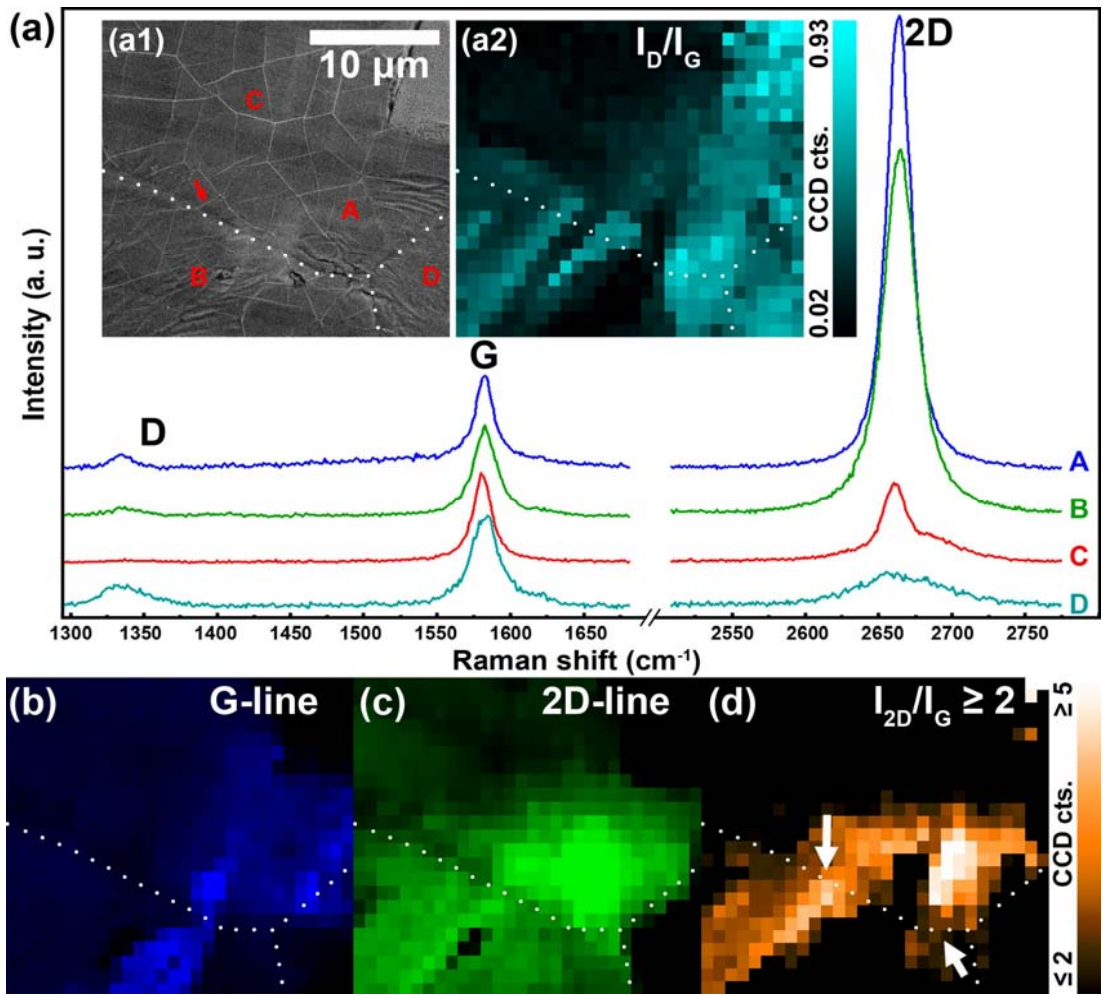

Fig. 1 - (a) Raman spectra from the positions A through D (marked in SEM image a1) of CVD synthesized graphene. The inset SEM image shows wrinkles of graphene as well as Ni grain boundaries (marked by red arrow and white dotted lines, respectively). (a2), (b), (c) and (d) are Raman images (corresponding to the SEM image) of $I_D/I_G$, G-line (1560 to 1620 cm$^{-1}$), 2D-line (2620 to 2720 cm$^{-1}$) and $I_{2D}/I_G \geq 2$, respectively.

The bridging of graphene wrinkles across surface features or grooves that are present on a substrate typically serves as indirect evidence that film growth bridges across grain boundaries. [13, 19, 24] In this report, the Raman images provide more direct evidence of the CVD graphene growth mechanism. Fig. 1a2 shows the Raman image obtained from an intensity ratio of D- and G-line ($I_D/I_G$), representing the degree of disorder of the graphite-based materials. [29] The areas of lighter contrast reveal the distribution of defects, indicating no obvious relation between Ni grain boundaries and defect distribution. Fig. 1b and c show the Raman images of the G- and 2D-line. Fig. 1d shows the image obtained when plotting an intensity ratio of $I_{2D}/I_G \geq 2$. Under the premise of the symmetrical shape of the 2D peak, it is generally acknowledged that the intensity of the 2D peak should be larger than G by over a factor of 2 for SLG and some areas of bi-layer graphene (BLG). [13, 21-23] With this in mind, Fig. 1d reveals SLG/BLG regions bridging across Ni grain boundaries (marked by white arrows in Fig. 1d). Although the distribution of graphene flakes is expected to be influenced by Ni grain boundaries during initial graphene precipitation, SEM and Raman analyses show no direct correlation between grain boundaries and graphene flakes.

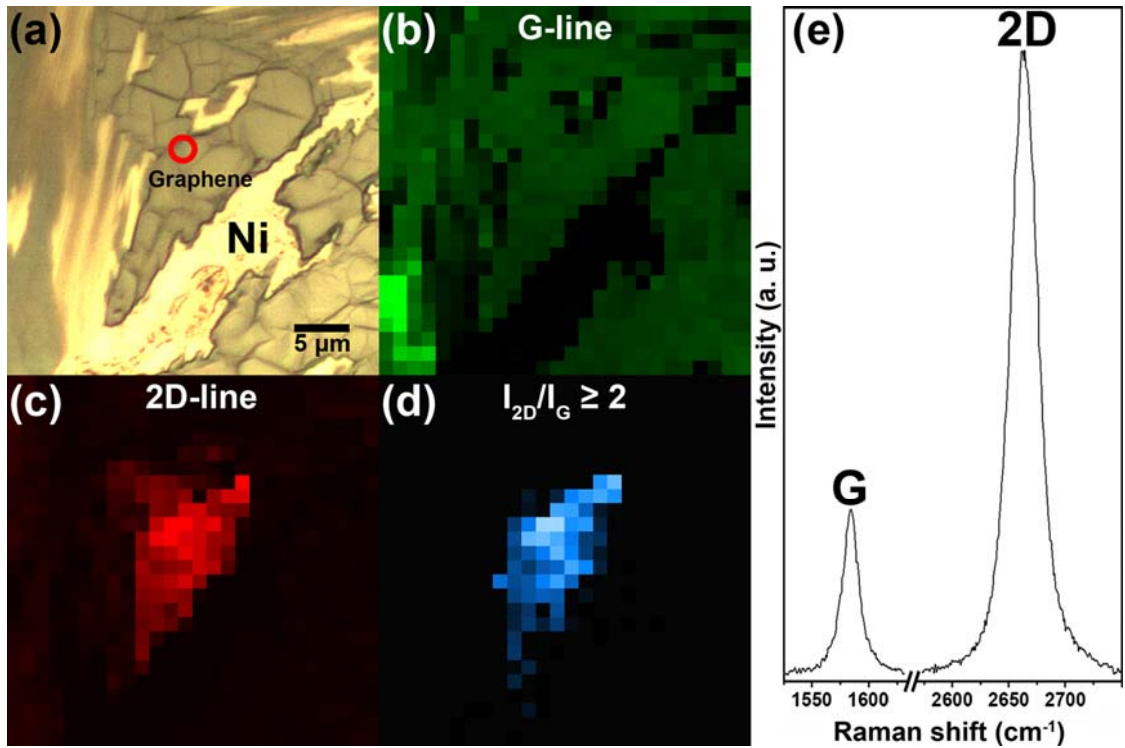

Fig. 2 - (a) to (d) OM and Raman images of as-grown graphene on Ni foil processed with a slow cooling rate of 0.3 ºC/sec. (e) Raman spectra taken at the position marked with a circle, indicating the presence of SLG to BLG.

Comparing the graphene samples fabricated using a slow and fast cooling rate (0.3 and 3 ºC/sec) in our CVD process, it is difficult to definitively extract the affect of cooling rate on the resulting film microstructure. Fig. 2 shows the Raman images and the SLG-to-BLG-featured Raman spectrum at the positions marked with circles in the OM image of the slow-cooled sample. The 2D peak can be fit with a single Lorentzian with full width at half maximum (FWHM) of 29 cm$^{-1}$. It is useful to consider Fick's laws of diffusion, which states that the diffusion path is proportional to $\sqrt{Dt}$, [30] where $D$ is diffusion coefficient and $t$ is time. A slower cooling rate provides more time for dissolved carbon atoms to migrate across the Ni, leading to more carbon atoms being precipitated on the Ni surface. According to this reasoning, the area of FLG should dramatically increase in samples subjected to the slow cooling process, leaving less SLG-to-BLG and empty area regions. However, the Raman images in Fig. 2 suggest the opposite. Fig. 2b and c show the G- and 2D-line images of Fig. 2a, and Fig. 2d shows the image of $I_{2D}/I_G \geq 2$. Areas of SLG-to-BLG and empty areas are seen to occupy a sizable fraction of the sample surface. After comparison to literature published on cooling-rate-controlled precipitation mechanisms in other systems, [12] this unexpected result suggests that the precise mechanism of surface precipitation/segregation will require additional experiments to verify.

It is known that carbon could be reduced from methane onto the Ni surface, similar to the CVD process used for the synthesis of carbon nanotubes. A possible explanation for our unexpected results, therefore, is that the deposition mechanism may play a larger role in final film formation than the precipitation mechanism. It is possible that both mechanisms occur simultaneously during graphene growth. This conclusion is similar to that of other reports that used Cu as the substrate, [24, 31] suggesting that the results they found for the Cu system could be extended to Ni substrates. Note that the graphene coverage in Fig. 2a does not represent the average coverage of entire sample surface. In fact, the visual coverage of fast and slow cooled samples are quite similar (>90 %).

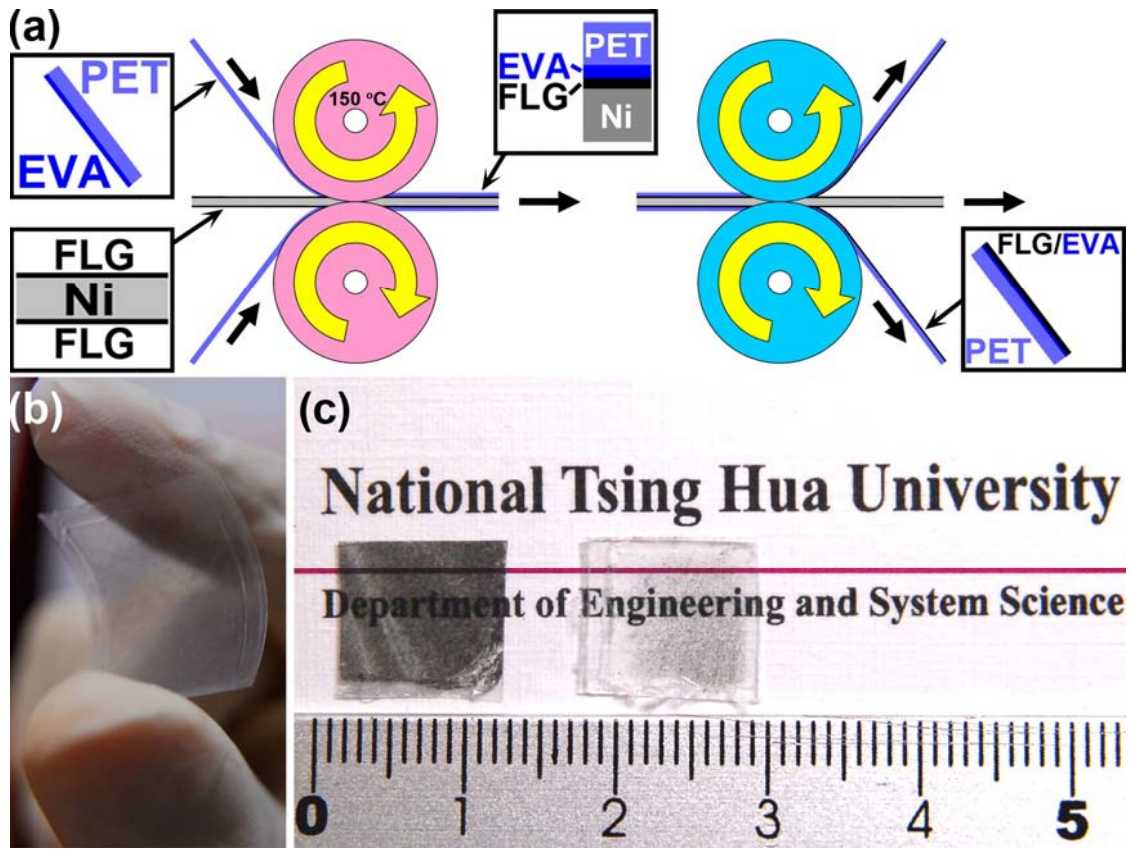

Fig. 3 - (a) A diagram to show the roll-to-roll process for the transfer of FLG from Ni foil to EVA/PET substrates. (b) A 2 × 3 cm$^2$ transparent flexible FLG/EVA/PET sample. (c) Two 1 × 1 cm$^2$ transferred samples with different transmittances.

Roll-to-roll processing is an ideal production choice when a very low cost per unit area of deposition is required. Fig. 3a is a diagram showing the roll-to-roll process for the transfer of FLG to various flexible substrates. In this process, commercial ethylene-vinyl acetate copolymer (EVA) coated transparent polyethylene terephthalate (PET) sheets were used as the target substrate. At a temperature of 150 °C, the EVA/PET and FLG/Ni sheets were pressed together with hot rollers to form a double-sided PET/EVA/FLG/Ni sheet, as shown in the inset in Fig. 3a. The EVA layer here plays a role of viscose between the PET and FLG. After the hot rolling step, the sheet was passed onto cold rollers at room temperature. The purpose of the cold rolling step is to separate the PET/EVA/FLG layers from the Ni surface in a uniform manner with a controlled, constant rolling speed. Most importantly, a large number of our experiments reveal that the thickness of FLG on PET is independent of rolling speed, and instead defined by the initial thickness of the FLG on the Ni surface. In other words, the transferred thickness of the graphene films could potentially be controlled according to the thickness of the deposited material during the CVD process. Fig. 3b is a 2 × 3 cm$^2$ transparent flexible FLG/EVA/PET sample produced using the roll-to-roll process. Samples of two different thicknesses are shown in Fig. 3c. The transmittance properties of the layered samples are inversely proportional to the thickness of the FLG layer; the sample shown on the left in Fig. 3c has a thicker FLG layer than the sample on the right.

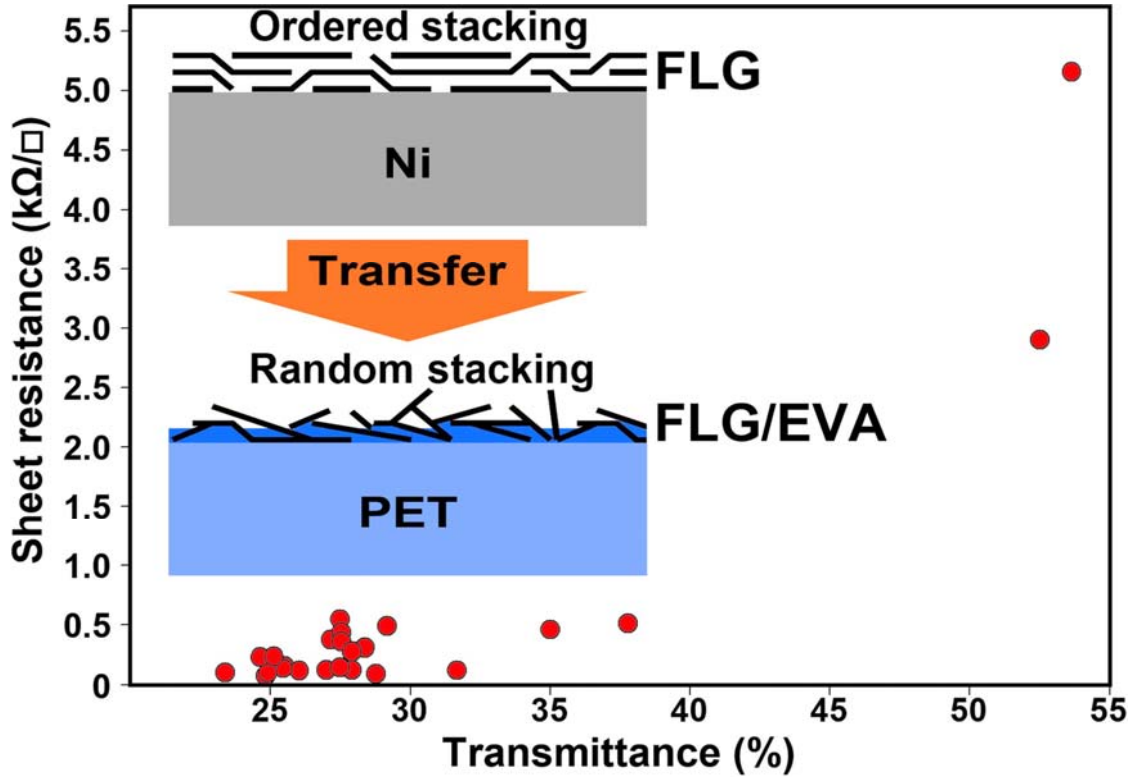

Fig. 4 - The plot of sheet resistance versus transmittance. The inset sketches the change of FLG stacking before and after roll-to-roll process.

Fig. 4 presents the plot of the sheet resistance versus transmittance of the FLG/EVA/PET samples. The results clearly reveal a positive correlation between sheet resistance and transmittance, as expected. To explain the undesirable property of high sheet resistance and low transmittance of the roll-to-roll transfer process, we suggest an explanation that is depicted in the inset of Fig. 4. Before transfer, the as-grown FLG on Ni surface has well-ordered stacking, but the FLG layer becomes more disordered after transfer. The random stacking results in poor contact between the FLG flakes, thus causing the increased sheet resistance. The increased roughness of the randomly-stacked FLG/EVA surface will also act as a scattering for the incident light, resulting in the decrease of transmittance. Future research on doping the graphene films to enhance sheet conductivity and planarizing the FLG/EVA surface for enhanced transmittance is necessary.

**4. Conclusions**

In conclusion, we have successfully synthesized SLG to FLG films using Ni foil as substrates via a CVD method. We point out that the growth method seen in Cu systems, [24, 31] namely the deposition mechanism, may be applicable to the Ni system in some cases. Furthermore, our experimental results suggest that controlling the cooling rate of the CVD process may not be an appropriate parameter to use for controlling the thickness of graphene films, namely because the deposition and precipitation mechanisms may occur simultaneously during graphene growth. In addition, we introduce an efficient way to transfer centimeter-scale FLG to transparent flexible PET substrates, and found that the thickness of the FLG on PET is dependent on the initial thickness of the FLG on the Ni surface. This report is the first time that a roll-to-roll process has been used for graphene transfer. After making a comprehensive survey of future graphene applications, it is obvious that this CVD and roll-to-roll process will be of great importance for the development of future graphene-based devices. Several industrial implications can be drawn from this study. First, in contrast to the thin film system, the metal foil is obviously less restrictive in terms of sample area. Second, compared to traditional rigid substrates, flexible metal foils allow for the use of a roll-to-roll transfer process. Third, due to undesirable high sheet resistance and low transmittance of these initial samples, future research should focus on doping the graphene films for increased sheet conductivity and planarizing the FLG/EVA surface to decrease light scattering. Finally, because the details of the growth mechanism of graphene in the CVD process remain uncertain, the application of advanced, nondestructive techniques for the analysis of as-grown graphene on the substrate interface would be very useful. We anticipate that the work we outline here will lead to important findings related to CVD and transfer processes for future applications.

**Acknowledgements**

The authors would like to thank Professor Po-Wen Chiu at the Institute of Electronics Engineering at National Tsing Hua University, Taiwan for fruitful discussions. This work was supported by the National Science Council of Taiwan under contract no. NSC 98-2218-E-007-003-MY3.